\documentclass[preprintnumbers,superscriptaddress,landscape,nofootinbib]{revtex4}
\usepackage[utf8]{inputenc}
\usepackage{amsmath}
\usepackage{amsthm}
\usepackage{amssymb}
\usepackage{float}
\usepackage{braket}
\usepackage{graphicx}
\usepackage{url}
\usepackage[colorlinks=true, pdfstartview=FitV, linkcolor=red, citecolor=blue, urlcolor=blue]{hyperref}
\usepackage{slashed}
\usepackage[normalem]{ulem}
\graphicspath{{./figures/}}

\newcommand{\be}{\begin{equation}}      
\newcommand{\ee}{\end{equation}}      
\newcommand{\bea}{\begin{eqnarray}}      
\newcommand{\eea}{\end{eqnarray}}

\newcommand{\ctext}[1]{\raise0.2ex\hbox{\textcircled{\scriptsize{#1}}}}

\usepackage{tikz}
\usetikzlibrary{quantikz}

\newtheorem{theorem}{Theorem}[section]

\newtheorem{conjecture}[theorem]{Conjecture}

\theoremstyle{definition}
\newtheorem{definition}[theorem]{Definition}

\theoremstyle{remark}

\begin{document}

\title{Quantum Extensive Form Games} 
\author{Kazuki Ikeda}
\email[]{kazuki7131@gmail.com}
\affiliation{Co-design Center for Quantum Advantage $\&$ Center for Nuclear Theory, Department of Physics and Astronomy, Stony Brook University, Stony Brook, New York 11794-3800, USA}
\affiliation{Department of Mathematics and Statistics
$\&$ Centre for Quantum Topology and Its Applications (quanTA), University of Saskatchewan, Saskatoon, Saskatchewan S7N 5E6, Canada}

\bibliographystyle{unsrt}

\begin{abstract}
Abstract: We propose a concept of quantum extensive-form games, which is a quantum extension of classical extensive-form games. Extensive-form games is a general concept of games such as Go, Shogi, and chess, which have triggered the recent AI revolution, and is the basis for many important game theoretic models in economics. Quantum transitions allow for pairwise annihilation of paths in the quantum game tree, resulting in a probability distribution that is more likely to produce a particular outcome. This is similar in principle to the mechanism of speed-up by quantum computation represented by Grover's algorithm. A quantum extensive-form game is also a generalization of quantum learning, including Quantum Generative Adversarial Networks. As an new example of quantum extensive-form games, we propose a quantum form of the Angel problem originally proposed by Conway in 1996. The classical problem has been solved but by quantizing it, the game becomes non-trivial. 
\end{abstract}

\maketitle

\section{Introduction}
Games are generally classified into two categories: those in which players act simultaneously and those in which they act sequentially. The games that triggered the AI revolution in recent years~\cite{silver2016mastering}, such as Go, Shogi, and chess, are what we call extensive-form games, which are applied to various foundations of economics~\cite{hart1992games}, computer information science and mathematical model of games. It is also possible to define a quantum version of the extensive-form games by applying a series of quantum operators to an initial quantum state. In fact several quantum versions of chess have been proposed~\cite{doi:10.1142/S0129626410000223,cantwell2019quantum}. The extensive-form is also very useful when discussing repeated non-cooperative games. In quantum games, single-stage games have long been favored~\cite{1999PhRvL..83.3077E,DBLP:journals/qip/KhanSBH18}, while games consisting of multiple stages have rarely been studied, although there are some exceptions~\cite{2007JMP....48g2107K,2022arXiv221102073I,ikeda2021quantum}. In our real world, however, we are far more likely to play games consisting of multiple stages than single-stage games. For example, we work in communities for long periods of time and buy and sell commodities, currencies, stocks, bounds and crypto assets every day. Considering that there are generally multiple processes involved in realistic negotiations and contracts, it is quite natural to study repeated or extensive-form games. Contrary to this circumstance, to the best of the author's knowledge, there are only a few examples of repeated quantum games being applied to economic models. In
\cite{ikeda_foundation_2019,ikeda_theory_2021}, the first infinite repeated quantum game was proposed, and it was applied to a macroeconomic model of quantum currency~\cite{2021QuIP...20..387I}.

In both classical and quantum games, the result of a game is defined by a probability distribution. The quantum extensive-form game proposed in this study is a general extension of what is called a mixed strategy type extensive-form game in the classical theory. While the strategy of a classical extensive-form game is given by a probability distribution in a real space, in a quantum extensive-form game the strategy is defined by an operator acting in a complex space. The difference between whether the space defining a stochastic transition is real or complex space is actually a very big difference. When transitions are defined in a real space, the probability distribution is always given by non-negative real numbers. On the other hand, if the transitions are defined in a complex space, the matrix components of strategy operators that yield the probability distribution are allowed to take negative values. This negative value allows pair annihilation of branches in the game tree. Therefore this makes it possible to manipulate the probability distribution so that a certain outcome is obtained with high probability. In the case of classical extensive form-games, such a pairing of branches of the game tree cannot occur since the parameters of the probability distribution cannot take negative values. In other words, once a probability distribution leading to a certain outcome is generated, it cannot be erased by any means. 

The idea of efficiently retrieving a particular value by pairwise annihilation of paths in the process of quantum state transformation is in fact the foundation of quantum algorithms. A typical example of an algorithm that achieves such quantum acceleration is Grover's algorithm, which solves the search problem of finding a specified value in an unordered database of $N$ elements~\cite{grover1998framework}.

\begin{figure}[H]
    \centering
    \includegraphics[width=8cm]{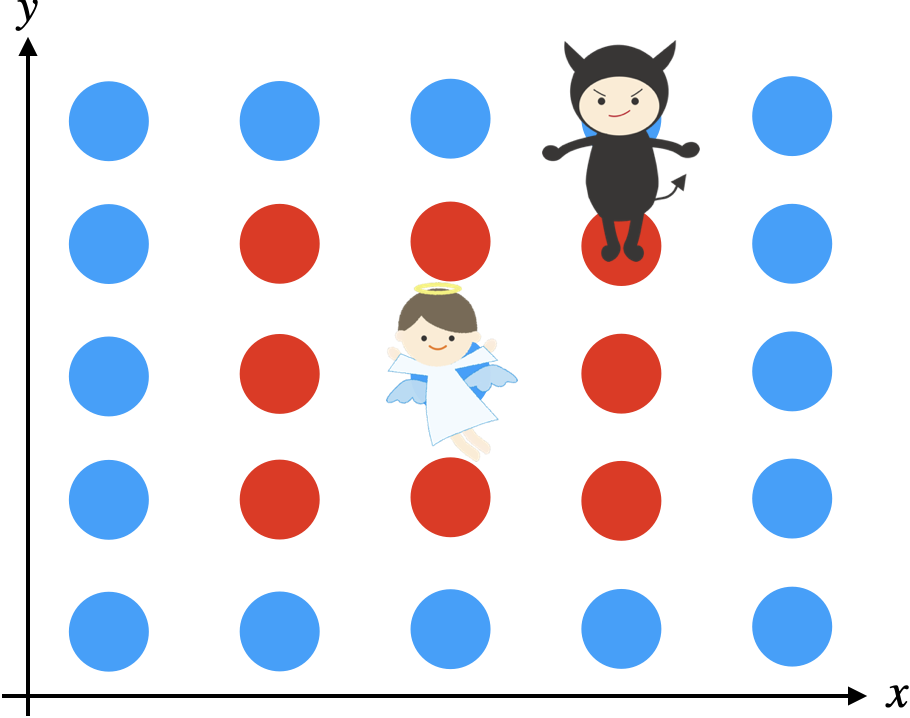}
    \caption{Conway's Angel Problem. In this picture, Devil succeeded in placing red blocks around Angel and was able to capture Angel.}
    \label{fig:AngelProblem}
\end{figure}

Angel problem is a extensive-form game proposed by Conway in 1996~
\cite{conway1996Angel}, where Angel and Devil take turns playing on a two dimensional lattice (Fig.~\ref{fig:AngelProblem}). Angel and Devil take turns playing the game. Devil places a block (red dots in the figure) on one lattice point each round to catch Angel. Angel runs away from Devil to avoid being surrounded by blocks. This game has been solved by some ways~\cite{mathe2007Angel,bowditch2007Angel,kloster2007solution}. In the traditional version, the status of the game is a common knowledge for all players. Such games are called games with complete information. In this paper, we quantize this game as follows. First, Angel moves on the graph as a quantum walker with power $k$. In other words, Angel moves in a quantum manner to a point within a distance $k$ from the point on the graph where it is currently located. In addition, the game itself consists of incomplete information, since the entire transition of the quantum state of Angel cannot be known. 

By quantizing angel problem, the diversity of the game can be realized in a variety of ways, giving it a complexity not found in the classical setting. Changing the topology or dimension of the graph defining the problem is one of the most natural extensions. In fact, such extensions can be found in the classical version. The quantum version, however, extends beyond mere geometric changes. The first case is when Angel and Devil have sufficient quantum resources, i.e., when they can all implement any quantum operation and perform all possible observations that are physically allowed. This setup may be the easiest to handle in obtaining analytical results for the game.

The second is when the quantum resources of Angel or (and) Devil are limited. This setup is most interesting in implementing and playing games with real quantum computers. This is because all the quantum computers we currently use are noisy and have a limitation in the shape of the graphs they can implement, so called NISQ devices~\cite{2018arXiv180100862P}. Thanks to noise, however, Angel and Devil each have ample opportunity to win. In the traditional angel problem, Angel has a winning strategy, but in the quantum version this is not always true and can be more exciting. In addition to the presence of noise, another reason why the solution to this game could be non-trivial is that Angel is a quantum walker with power $k$, whose behavior will be extremely hard to predict. Extending the quantum computing power of Angel and Demon can also make the game more interesting. Computational theory based on theories beyond the quantum theory of reality is of great importance to the field of computer science and has been the subject of much research~\cite{2004quant.ph..1062A,aaronson2009closed}. 

Asking to what extent the quantum computational power of Angel and Devil can be limited or extended to make the game more satisfactory for each would be a useful new question in mechanism design as well as the computational complexity. For example, if the game can be designed so that Angel and Devil play the game enough times and each has a 50\% chance of winning, then it is fair game for both sides. While the conventional studies of quantum games have so far focused on the strategic aspects not found in classical games, there has been little research on optimal designs of game environments for players. In traditional game theory, designing optimal rules and games is practically very important~\cite{10.2307/1817047,10.2307/2566947,doi:10.1287/moor.6.1.58}. Looking at the quantum angel problem from a viewpoint of mechanism design, we can ask the following question. Angel and Devil need to rent quantum computing resources from one company, and the fee depends on the performance of the quantum computers available. They want to win the game as often as possible and generate as much revenue as possible on a limited budget. Given the respective budgets of Angel and Devil, how can we design the game so that their expected revenues can be fairly distributed? Such a general problem of optimal allocation of quantum resources is likely to arise in the near future.

The main results of this paper and its contributions to quantum game theory can be summarized as follows.
\begin{itemize}
    \item We gave a formal and rigorous formulation of quantum extensive-form games.
    \item We explained the importance of quantum quantum extensive-form games in relation to algorithms with quantum advantage.
    \item {We defined the Nash equilibrium of a general quantum extensive form game and discussed how quantum advantage can be realized.}
    \item We proposed the quantum angel problem as a new problem of computational complexity and mechanism design.
\end{itemize}

This paper is organized as follows. First, we will elaborate on the significance and motivation of quantum extensive-form games and establish a formal definition. Next, we formulate the quantum angel problem in concrete terms and give the quantum operators to implement it. Finally, we describe the role of this research in the development of game theory and suggest various future directions.

\section{Formulation quantum extensive-form games and Quantum Advantages}
\subsection{\label{sec:ad}Quantum Advantages in Repeated Games and Extensive Games}
Extensive-form quantum games give very interesting results not found in classical games. To discuss why quantum extensions of extensive form games and repeated games are effective, let us first discuss how quantum speedups of computation can be achieved by quantum algorithms. We consider the simplest repeated game consisting of two stages, as illustrated in the figure below. 
\begin{figure}[H]
    \centering
    \includegraphics[width=15cm]{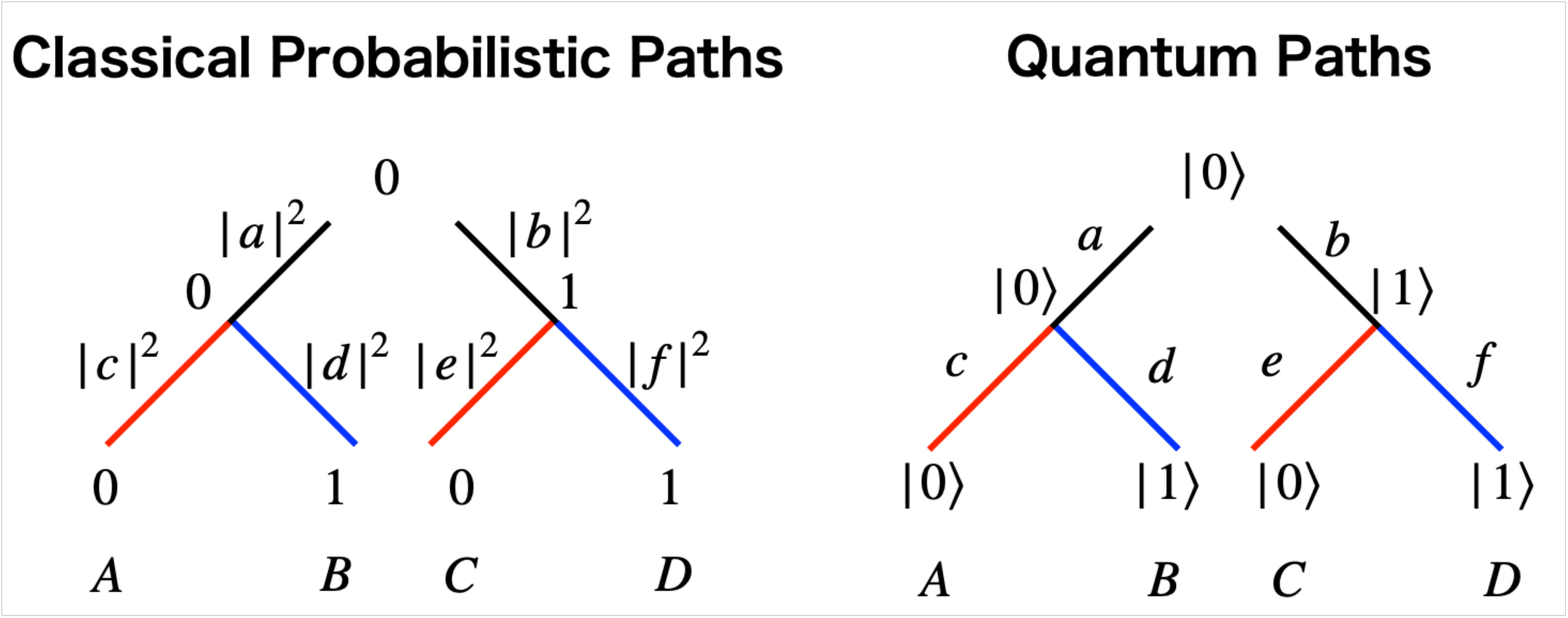}
    \caption{Transitions of states in classical and quantum game trees. Pair annihilation of branches can happen in a quantum game tree.}
    \label{fig:my_label}
\end{figure}

The game is played by outputting 0 or 1, and some reward is given based on the result.  Assuming the initial bit is 0, we consider paths that move to 0 or 1 with a non-zero probability $(a,\cdots,f\neq0)$. By the conservation law of probability, the parameters $a,\cdots,f$ satisfy $|a|^2+|b|^2=1$, $|c|^2+|d|^2=1$ and $|e|^2+|f|^2=1$. First, the classical probabilities of obtaining 0 or 1 in the first step are $|a|^2,|b|^2$ respectively:
\begin{align}
\begin{aligned}
\text{Prob}_\text{classical}(0,t=1)&=|a|^2>0\\
\text{Prob}_\text{classical}(1,t=1)&=|b|^2>0
\end{aligned}
\end{align}

The probability of moving to 0 or 1 after two steps is given by
\begin{align}
\begin{aligned}
\text{Prob}_\text{classical}(0,t=2)&=A+C=|a|^2|c|^2+|b|^2|e|^2>0\\\text{Prob}_\text{classical}(1,t=2)&=B+D=|a|^2|d|^2+|b|^2|f|^2>0
\end{aligned}
\end{align}
It is obvious, unless $a,\cdots,f$ are zero, these probabilities will never be zero. 

However, an interesting thing happens in the case of quantum stochastic transitions. To discuss this point, let us consider the case where the unitary transition is given by the following matrices: 
\begin{align}
\begin{aligned}
U_1&=a|0\rangle\langle0|+b|1\rangle\langle0|+a'|0\rangle\langle1|+b'|1\rangle\langle1|\\&=\begin{pmatrix}a&a'\\b&b'\end{pmatrix}\\
U_2&=c|0\rangle\langle0|+d|1\rangle\langle0|+e|0\rangle\langle1|+f|1\rangle\langle1|\\&=\begin{pmatrix}c&e\\d&f\end{pmatrix}.
\end{aligned}
\end{align}

As before, we assume that the parameters $a,\cdots,f$ are non-zero. Since we are not interested in transitions from $\ket{1}$, $a'$ and $b'$ can be anything that satisfies the unitary condition.
First, the probabilities of obtaining 0 or 1 in the first step after applying $U_1$ are
\begin{align}
\begin{aligned}
\text{Prob}_\text{quantum}(0,t=1)&=|a|^2>0\\
\text{Prob}_\text{quantum}(1,t=1)&=|b|^2>0
\end{aligned}
\end{align}

At this point, there is no difference from the case of the classic game. The probability of starting from state $\ket{0}$ and transitioning to state $\ket{k}~(k=0,1)$ after two steps is generally given by the following formula
\begin{equation}
    \text{Prob}_\text{quantum}(k,t=2)=|\langle i|U_2U_1|0\rangle|^2.
\end{equation}

Therefore the probabilities of transitioning from $\ket{0}$ to $\ket{1}$ two steps later are
\begin{align}
\begin{aligned}
\text{Prob}_\text{quantum}(0,t=2)&=|A+C|^2=|ac+be|^2\ge0\\
\text{Prob}_\text{quantum}(1,t=2)&=|B+D|^2=|ad+bf|^2\ge0
\end{aligned}
\end{align}

What is strikingly different from the classical case is that even if the parameters are not zero, the probability of 0 being output after two steps can be zero if $ac+be=0$. In this case, the probability of 1 being output is 1 
\begin{equation}
    \text{Prob}_\text{quantum}(0,t=2)=0\Leftrightarrow\text{Prob}_\text{quantum}(1,t=2)=1.
\end{equation}

The most important point is that such $U_2$ always exists, regardless of the transition in the first step. The simplest candidate for such $U_2$ is $U^\dagger_1$, which does exist since $U_1$ is unitary. The reason this cannot be achieved with classical stochastic transitions is that the probability parameters of classical stochastic transitions are always positive and the transitions are irreversible. On the other hand, quantum mechanical transitions are unitary and the parameters are allowed to be negative, which essentially contributes to creating such a quantum advantage. It is sufficient that negative numbers are permitted in the unitary transition matrices.

The operation of applying an operator two or more times $U_2U_1\ket{0}$ is truly an essence of repeated quantum games, and by strategically adjusting the probability density distribution, it is possible to achieve an advantageous result that would never be obtained in a classical game or a single-stage game. This also works well even for finitely repeated quantum games. This is in contrast to the classical repeated games, where a finite number of iterations can be considered equivalent to a single-stage game.

To make the above more game-theoretic, let $g_i(k)$ be the reward for Player i when outcome $k\in\{0,1\}$ is obtained. The player applies sequentially unitary operators $U_1,\cdots,U_k$ to the initial state $\ket{0}$. The expected profit when the state $\ket{0}$ is observed at time $t$ is given by
\begin{equation}
\label{eq:profit}
    g_i\left(t,\{U_{\tau}\}_{\tau=1}^t\right)=\sum_{k\in\{0,1\}}g_i(k)|\langle i|U_{t}U_{t-1}\cdots U_1|0\rangle|^2.
\end{equation}

The discussion and observations in this section can be summarized as follows.
\begin{itemize}
    \item Certain outcomes can appear or disappear due to pairwise annihilation of certain paths. By annihilating paths that lead to unfavorable outcomes with a high probability of success, the desired outcome is more likely to occur. This means, for example, that Pareto-optimal or better solutions are more likely to be realized.
    \item In the case of quantum games, even a finitely repeated game is not equivalent to a single-stage game. A finitely repeated quantum game can also produce superior results that cannot be obtained in a classical single-stage game. This is because of a pair annihilation of paths during a unitary time-evolution of a game, which never happens in any finitely repeated classical game.
    \item Repeated quantum games might look like a quantum extension of stochastic game, where time-evolution is given by classical probabilistic transitions. However once a classical path is generated by classical probabilistic transitions, then the path never vanishes. Therefore, in quantum games, when path annihilation occurs during the unitary time evolution of a game, we can find quantum supremacy.
    \item These ideas are valid not only for repeated games, but also for games with several different stages and extensive-form games. For example, some effective results showing quantum advantages in a game theory of contracts were first reported in~\cite{ikeda2021quantum} for quantum principal-agent problems, adverse selections and moral hazard problems.
\end{itemize}

\subsection{Quantum Game Tree and Strategy}
Now let us give a formal definition of quantum extensive-form games. 
\begin{definition}
An quantum extensive-form game is defined by a tuple $\Gamma^Q=(N,K,P,I,g)$ consisting of the following date:
\begin{itemize}
    \item $N=\{1,\cdots,n\}$: A set of $n$ players.
    \item A quantum game tree $K$ with a set of quantum states of nodes $V$, a initial quantum state of node $v_0\in V$. Each node can be classified into either a vertex or a move: $V=X\cup W,X\cap W=\emptyset$, where $X$ is the set of all moves and $W$ is the set of all vertexes. 
    \item Player partition $P=\{P_1,\cdots,P_n\}$ that assigns each player to a set of moves: $X=\bigcup_{i=1}^nP_i$, $P_i\cap P_j=\emptyset$ for all $i,j$ and $P_i\neq\emptyset$. 
    \item Information partition $I=\{I_1,\cdots,I_n\}$. Each $I_i$ is a family of subsets of $P_i$, which is written as $P_i=\bigcup_{u\in I_i}u$. Elements of $I_i$ satisfy $u\cap v=\emptyset$ for all $u,v\in P_i$ such that $u\neq v$. 
    \item Payoff function $g(w)=(g_1(w),\cdots,g_n(w)),\ket{w}\in W$ of all players.  
\end{itemize}
\end{definition}

A quantum game is called a quantum game with complete information when everyone knows the state of the game and the possible moves of all other participants, otherwise it is a quantum game with incomplete information. In many cases, quantum extensive form games consists of imperfect information, since the quantum states that make up the game cannot be known exactly.

Let $B(\ket{\psi})$ be the set of quantum moves connected by alternatives from $\ket{\psi}$. In other words, $B(\ket{\psi})$ is the set of points immediately after a quantum move $\ket{\psi}$. We define the set of quantum moves from $u\in I_i$ by 
\begin{equation}
B(u)=\bigcup_{\ket{\psi}\in u}B(\ket{\psi}).    
\end{equation}
In a classical extensive form game, a classical move $x$ and each point of $B(x)$ are connected by a branch called alternative. The set of such alternatives is described as $A(x)$ and $A(u)=\cup_{x\in u}A(x)$ is the set of alternatives in the information set $u\in I_i$. A strategy in a classical extensive form game is given by a map from $u\in I_i$ to $A(u)$. In the quantum extensive form games, a strategy is a map from a quantum state to quantum state. 

\begin{definition}
A quantum strategy of Player $i$ is a quantum channel $\Lambda$ from $u\in I_i$ to $B(u)$. 
\end{definition}
A map $\Lambda$ that transfers a quantum state $\ket{\psi}$ in a Hilbert space $\mathcal{H}$ to another quantum state $\ket{\psi'}=\Lambda(\ket{\psi})$ is called a quantum channel. The Hilbert space to which the quantum state $\ket{\psi}$ is mapped may be different from the original space $\mathcal{H}$. It can be the same Hilbert space if all players are playing the game on the same platform, or a different Hilbert space if they are using different platforms. For example, a quantum crypto key is defined as a map from a high-dimensional Hilbert space to a low-dimensional Hilbert space.

\begin{figure}[H]
    \centering
    \includegraphics[width=15cm]{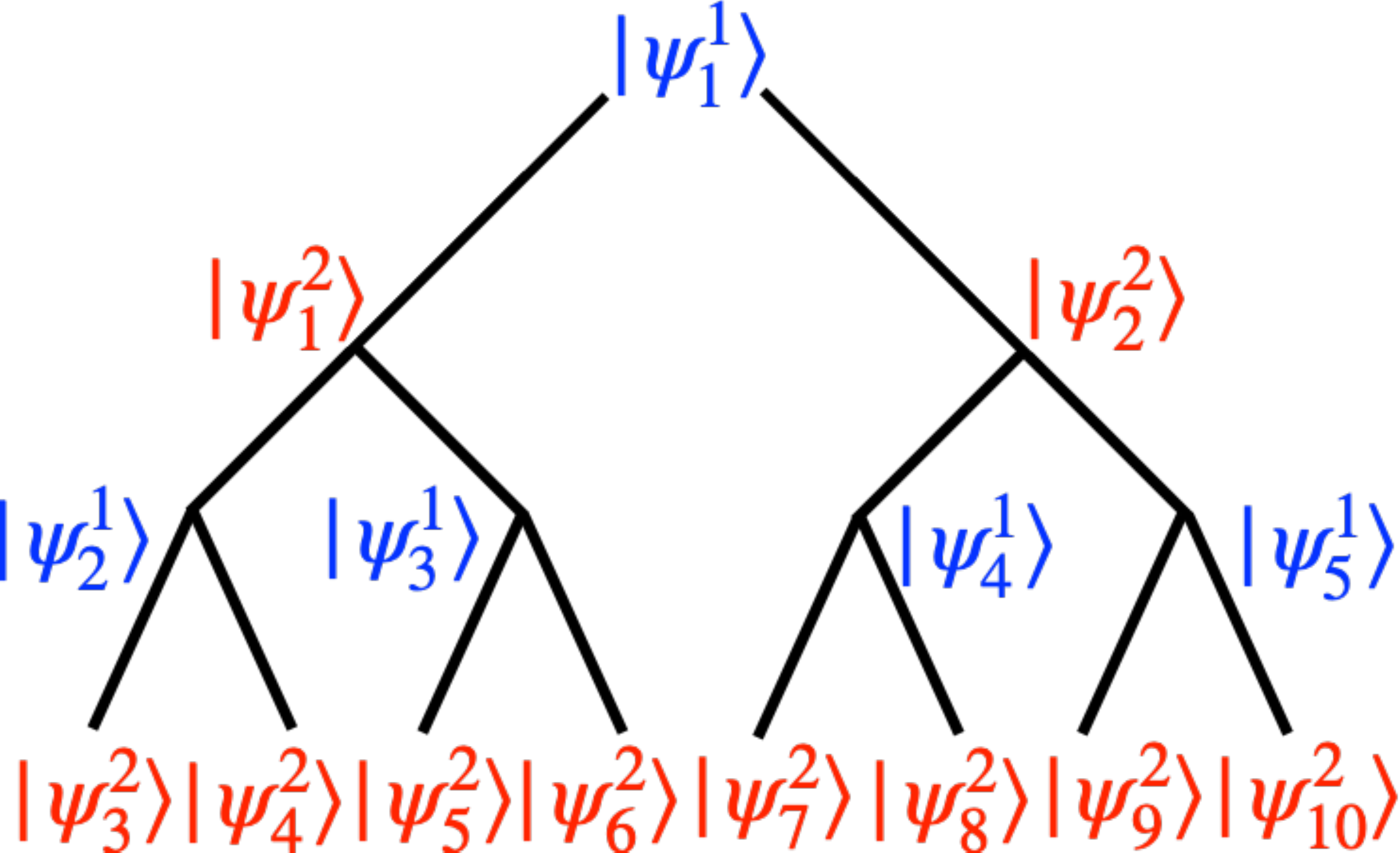}
    \caption{Quantum game tree of an quantum extensive-form game.}
    \label{fig:extensive}
\end{figure}

Let us illustrate the structure of a quantum extensive-form game using the example in Fig.\ref{fig:extensive}. In this game two players take turns playing. 

In this setup, the player partitions are 
\begin{align}
    \begin{aligned}
    P_1&=\{\ket{\psi^1_1},\cdots, \ket{\psi^1_5}\}\\
    P_2&=\{\ket{\psi^2_1},\cdots, \ket{\psi^2_{10}}\}.
    \end{aligned}
\end{align}
Information partitions could be 
\begin{align}
    \begin{aligned}
    I_1&=\{\{\ket{\psi^1_1}\},\{|\psi^1_2\rangle,|\psi^1_5\rangle\}, \{|\psi^1_3\rangle,|\psi^1_4\rangle\}\}\\
    I_2&=\{\{\ket{\psi^2_1},\ket{\psi^2_2}\},\{|\psi^2_3\rangle,\cdots,|\psi^2_6\rangle\}, \{|\psi^2_7\rangle,\cdots,|\psi^2_{10}\rangle\}\}.
    \end{aligned}
\end{align}
There can be many other ways to choose information partitions as long as they satisfy $P_i=\bigcup_{u\in I_i}u$ and $u\cap v=\emptyset$ for all $u,v\in P_i$ such that $u\neq v$. 

For example, the sets of quantum moves connected by alternatives from $\ket{\psi^1_1}$ and $\ket{\psi^2_1}$ are
\begin{align}
    \begin{aligned}
    B(\ket{\psi^1_1})&=\{\ket{\psi^2_1},\ket{\psi^2_2}\}\\
    B(\ket{\psi^2_1})&=\{\ket{\psi^1_2},\ket{\psi^1_3}\}. 
    \end{aligned}
\end{align}    
Therefore when $u=\{\ket{\psi^2_1},\ket{\psi^2_2}\}\in I_2$, then the corresponding set $B(u)$ of quantum moves is 
\begin{equation}
    B(u)=\{\ket{\psi^1_2},\ket{\psi^1_3},\ket{\psi^1_4},\ket{\psi^1_5}\}.
\end{equation}
Given the sets of quantum moves $u$ and $B(u)$ in this way, a quantum strategy of Player 2 is a quantum channel from $u$ to $B(u)$. The Hilbert space in which $u$ is defined and the Hilbert space in which $B(u)$ is defined may be different in general. This transfers Player 2's turn to Player 1, who then considers a new quantum strategy.

\subsection{Best Response and Nash Equilibrium}
{In what follows we discuss players' strategies and the Nash equilibrium solutions. To make the games concrete, we address the case where quantum process is unitary. Therefore all quantum strategies are given by unitary matrices. }

{Let $S_i$ be the set of all quantum strategies of Player $i$ and $S_{-i}$ be the set of all quantum strategies for everyone except Player $i$. Let $U_{-i}\in S_{-i}$ denote a quantum strategy (unitary operator) of everyone except Player $i$. }
{
\begin{definition}
When a quantum game tree $\Gamma^Q$ and sets $S_i,S_{-i}~(i=1,\cdots,N)$ of strategies of $N$ players are given, we call $U^*_i\in S_i$ a best response against $U_{-i}\in S_{-i}$ if it satisfies the following condition: 
\begin{equation}
    g_i(U^*_i,U_{-i})\ge g_i(U_i,U_{-i}),~\forall U_i\in S_i,
\end{equation}
where $g_i$ is the profit function of Player $i$.  
\end{definition}}

{It is important to note that there may be more than one best response. For example, there are innumerable best responses that give the same result, differing only in the phase factor of a quantum state.
\begin{definition}
For a given quantum strategy $U_{-i}\in S_{-i}$ of everyone except Player $i$, let 
    \begin{equation}
        b_i(U_{-i})=\{U^*_i\in S_i: g_i(U^*_i,U_{-i})\ge g_i(U_i,U_{-i}),~\forall U_i\in S_i \}
    \end{equation}
be the set of all best responses of Player $i$. We call this assignment 
\begin{equation}
    b_i:S_{-i}\to b_i(S_{-i})\subset S_i
\end{equation}
the best response correspondence. 
\end{definition}
}

{
\begin{definition}
A strategy set $(U^*_1,\cdots, U^*_N)$ of $N$ players are called Nash equilibrium if each strategy $U^*_i~(i=1,\cdots, N)$ is a best response to the strategies $U^*_{-i}=(U^*_1,\cdots, U^*_N)\setminus{U^*_i}$ of the other players. In other words, a Nash equilibrium strategy of Player $i$ satisfies
\begin{equation}
    \forall i\in\{1,2,\cdots,N\},~\forall U_i\in S_i,~g_i(U^*_i,U^*_{-i})\ge g_i(U_i,U^*_{-i}).
\end{equation}
\end{definition}
}

{What Nash equilibrium means is that, when all other players except Player $i$ are playing the game in accordance with the equilibrium strategy, Player $i$ does not gain any advantage by deviating from the equilibrium. In the case of quantum unfolding games, the Nash equilibrium may be very different from the classical case and may not always exist. In some specific examples, Nash equilibria have been considered. For example, in the quantum version of the ultimatum game, which is a typical example of an extensive form game, it has been reported that a Nash equilibrium exists that is different from the classical equilibrium~\cite{mendes2005quantum}.}

{
\begin{definition}
    Let $\Gamma^Q$ be a quantum extensive form game and $\ket{\psi}$ be a quantum state that is a node of the game tree. Then we denote by $P_\text{quantum}(\psi|U_I)$ the probability of reaching state $\ket{\psi}$ with a set $U_I=(U_1,\cdots,U_N)$ of quantum strategies of $N$-players. We call a state $\ket{\psi}$ reachable if there exists such a $U_I$ that 
    \begin{equation}
        P_\text{quantum}(\psi|U_I)>0.
    \end{equation}
\end{definition}
}

{In the quantum extensive form game, the most striking difference from the classical case is the reachability of certain nodes. This probability is nothing but the probability amplitude from the initial node to another node $\ket{\psi}$
\begin{equation}
    P_\text{quantum}(\psi|U_I)=|\bra{\psi}U_NU_{N-1}\cdots U_1\ket{\psi_{in}}|,
\end{equation}
where $\ket{\psi_{in}}$ is the initial quantum state. This exactly corresponds to that given on the right-hand side of equation~\eqref{eq:profit}. In the classical case, it is obtained by sequentially multiplying the probability of reaching each node on the branch on which that node is reached. It is this difference that makes the quantum extensive form game different against the classical, as discussed in Section~\ref{sec:ad}, and in some cases realizes a quantum advantage, such as Grover's algorithm and Shor's algorithm. In the classical game, the probability of reaching a node is zero if and only if the probability of reaching a node on the branch leading to that node is zero.}

{
\begin{definition}
    Let $\Gamma^Q$ be a quantum extensive form game and $K$ be the quantum game tree. Let $K'$ be a sub-tree of $K$ such that the restriction of the information set of the original game $\Gamma^Q$ on $K'$ never contains the moves of $K'$ and the moves other than $K'$ at the same time. Then the restriction of $\Gamma^Q$ to $K'$ can be regarded as a game with a game tree $K'$, which we call a subgame.   
\end{definition}
Let $W^\psi$ be the set of all vertices in the subgame $\Gamma^Q(\psi)$. Then the expected payoff of Player $i$ when a set $U_I(\psi)$ of strategies are played is given by  
\begin{equation}
    f^\psi_i(U_I(\psi))=\sum_{k\in W^\psi}P_\text{quantum}(k|U(\psi))g_i(k)
\end{equation}
}

{
\begin{definition}
Let $\Gamma^Q$ be a quantum extensive form game, and $\Gamma^Q(\psi)$ be a subgame with an initial quantum node $\ket{\psi}$ and $U_I(\psi)=(U_1,\cdots,U_N)$ be a set of strategies on a subgame $\Gamma^Q(\psi)$. A game that can be made by taking the entire subgame as one vertex of the original game $\Gamma^Q$ and replacing the profit at that point with the expected profit calculated by $U_I=(U_1,\cdots,U_N)$ is called a truncated game $T(\Gamma^Q/\Gamma^Q(\psi),U_I(\psi))$.  
\end{definition}
}

{The following theorem can be shown immediately by proof by contradiction.
\begin{theorem}
A set $U^*=(U^*_1,\cdots,U^*_N)$ of Nash equilibrium strategies of a quantum extensive form game $\Gamma^Q$ satisfy the following two properties:
\begin{enumerate}
    \item It gives a set $U'^*$ of equilibrium strategies in all subgames reachable by $U^*$.
    \item For any subgame $\Gamma^Q(\psi)$, it gives a set $U''^*$ of equilibrium strategies in the truncated game $T(\Gamma^Q/\Gamma^Q(\psi),U_I(\psi))$.
\end{enumerate}
\end{theorem}
The equilibrium solution of a subgame that is given by the property 1 of the theorem is what is called a subgame perfect equilibrium in conventional games. Note that A subgame perfect equilibrium is a Nash equilibrium, but a Nash equilibrium does not necessarily coincide with a subgame perfect equilibrium. This is the same as in the classical case. A feature of quantum extensive form games is the possibility of obtaining a Nash equilibrium that differs from the classical case. Moreover as we frequently discuss, it may converge to an} equilibrium solution faster than the classical extensive form games, due to pair annihilation of paths.

\subsection{Relation to Former Works}
In order to most clearly convey the significance of the quantum extensive-form game proposed in this study, we first compare it with Grover's algorithm~\cite{grover1998framework}, which is a well-known quantum algorithm showing quantum speedup. The goal of this algorithm is to efficiently find the desired information $w$ from a database consisting of $N$ data points. The initial state is the uniform superposition of all data
\begin{equation}
    \ket{\psi_\text{in}}=\frac{1}{\sqrt{N}}\sum_{i=0}^{N-1}\ket{i}. 
\end{equation}
The quantum operator $U_2U_1$ is applied on this initial state, where
\begin{align}
    U_1&=2\ket{\psi_\text{in}}\bra{\psi_\text{in}}-I\\
    U_2&=I-2\ket{w}\bra{w}. 
\end{align}
It is truly an extensive-form game in the sense that $U_1$ and $U_2$ are applied multiple times in turn, and transitions of the initial quantum state can also be described using a game tree. w can be found by iteratively applying the Grover iteration $U_1U_2$ about $O(\sqrt{N})$ times, which is known to take $O(N)$ times in the classical best way. This is made possible by efficiently eliminating branches in which anything other than $w$ appears.

An important example of quantum extensive-form games is quantum generative adversarial learning (GAN) \cite{PhysRevLett.121.040502,zoufal2019quantum}, which increases the accuracy of data learning by allowing two neural networks to compete with each other. The most obvious example would be the emergence of quantum GANs for playing quantum chess, but of course, the applications are not limited to such simple examples. Even classical GANs are used everywhere, including image and video generation, and it is not difficult to foresee that in the future quantum GANs will play an equally or even more important role than current classical GANs. Any such quantum GAN can always be expressed in a quantum extensive-form.

Although it has not been given much importance in the past, quantum extensive-form games are a new way of thinking that can handle several important examples like this in a general way. By establishing a formal theory of quantum extensive-form games, it will be possible to conduct research on quantum algorithms from a bird's-eye, multifaceted, and unified perspective. {In fact, it is a very powerful method for game-theoretic analysis of events that occur in real markets. It has traditionally been a very useful method for examining market oligopolies and monopolies, and ultimatum games and repeated games are also typical examples of extensive form games. }

\section{Quantum Angel Problem}
\subsection{Overview}
\begin{figure}[H]
    \centering
    \includegraphics[width=8cm]{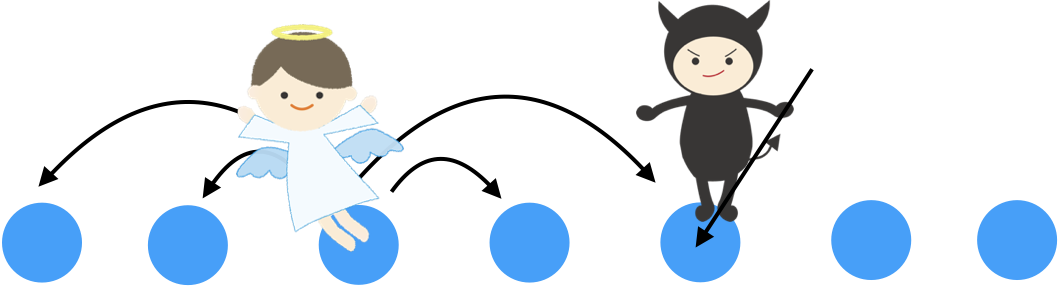}
    \caption{Quantum Angel on a one dimensional chain of qubits. Devil measures qubits to place blocks and catch Angel.}
    \label{fig:QAP}
\end{figure}

As an example of quantum extensive-form games, let us create a quantum version of angel problem. The classical angel problem was invented by Conway in 1996~\cite{conway1996Angel} and investigated by many authors~\cite{mathe2007Angel,bowditch2007Angel,kloster2007solution}. The quantum angel problem is played by two players called Angel and Devil~(Fig.~\ref{fig:QAP}). Angel is a quantum walker on a given graph $G$ and move on the graph within the range allowed by the predetermined power $k$. Devil can add blocks to one vertex of the graph where there is no Angel. It is easy to extend this game so that Devil puts blocks (quantum) statistically. 

The rules are basically the same as before, except that Angel is a quantum walker. Angel and Devil take turns playing their respective strategies. Devil wins if Angel is unable to move, and Angel wins if she can keep running from Devil's pursuit. Angel can quantum tunnel through the the blocks, but cannot be in the same position as the blocks, in accordance with Pauli's exclusionary rule.

\begin{itemize}
    \item Angel can move to any place within a distance $k$ where no blocks are placed.
    \item Devil can place a block if no Angel is detected at the location where they want to place the block.
\end{itemize}

\subsection{Quantum Angel with Power $k$}
\begin{figure}
  \begin{center}
    \begin{tikzpicture}[node/.style={draw=blue, fill = orange!30, circle, minimum size=1cm}]
      \node[node, line width=0.3mm] (s1) {$x$};
      \node[node, right = 1cm and 1cm of s1, line width=0.3mm] (s2) {$x+1$};
      \node[node, left  = 1cm and 1cm of s1, line width=0.3mm] (s3) {$x-1$};
      \node[node, right  = 1cm and 1cm of s2, line width=0.3mm] (s4) {$x+2$};
      \node[node, left  = 1cm and 1cm of s3, line width=0.3mm] (s5) {$x-2$};
      \node[node, right  = 1cm and 1cm of s4, line width=0.3mm] (s6) {$x+3$};
      \node[node, left  = 1cm and 1cm of s5, line width=0.3mm] (s7) {$x-3$};
      \path[->,thick, blue, >=stealth]
        (s1) edge[left, bend right=20] node[anchor=north]{$U_1$} (s2)
        (s1) edge[left, bend left=20] node[anchor=north]{$U_{-1}$} (s3)
        (s1) edge[loop above] node[anchor=south]{$U_0$} (s1)
        (s1) edge[right, bend left=25] node[anchor=south]{$U_2$} (s4)
        (s1) edge[left, bend left=35] node[anchor=south]{$U_3$} (s6)
        (s1) edge[left, bend right=25] node[anchor=south]{$U_{-2}$} (s5)
        (s1) edge[left, bend right=35] node[anchor=south]{$U_{-3}$} (s7);
        \path[-,thick, >=stealth]
        (s7) edge[below]  (s5)
        (s5) edge[below]  (s3)
        (s3) edge[below] (s1)
        (s1) edge[below]  (s2)
        (s2) edge[below]  (s4)
        (s4) edge[below]  (s6);
    \end{tikzpicture}
  \end{center}
    \caption{Hopping of a quantum walker with power $k$.}
    \label{fig:QW}
\end{figure}
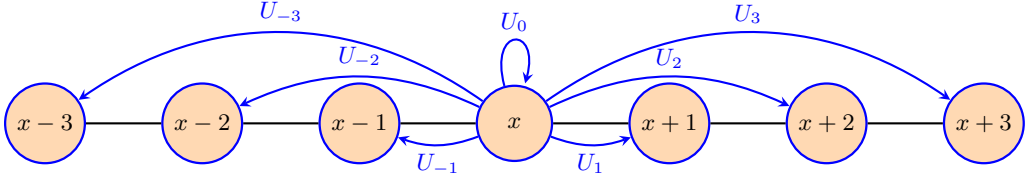

Here we formulate the general behavior of Angel with power $k$. For this, we first introduce a quantum walker that can move to a point within a range of at most $k$~(Fig.~\ref{fig:QW}). The position of Angel is described by a Hilbert space $\mathcal{H}_P$ spanned by integer values on the lattice $\{\ket{i},i\in \mathbb{Z}\}$. Quantum walkers that can travel up to distance $k$ in one time step do not seem to have been studied much. 

In this paper, for the sake of brevity and concreteness, the discussion will be limited to a one-dimensional lattice, but it can be extended to arbitrary graphs in any dimension, which would be a meaningful question. The most general form of Angel's state at $t=n$ is given below:
\begin{equation}
    \ket{\Psi_n(x)}=\sum_{m=-k}^k\psi^m_n(x)\ket{m},~\sum_{m=-k}^k|\psi^m_n(x)|^2=1
\end{equation}
where $x\in\mathbb{Z}$ indicates the position of Angel on a 1D lattice. The probability measure of Angel at time $t=n$ at location $x$ is defined by
\begin{equation}
    \mu_n(x)=\sum_{m=-k}^k|\psi_n^m(x)|^2. 
\end{equation}

The transition operator which maps Angel to $\ket{m}$ is given by 
\begin{equation}
    U_m=\sum_{l=-k}^ka_{lm}\ket{m}\bra{l},
\end{equation}
where the condition of the matrix $a=(a_{lm})$ is determined by the unitarity of $U$
\begin{equation}
    U=\sum_{m=-k}^kU_m,~U^\dagger U=1. 
\end{equation}
It should be noticed that each operator does not commute in general, which allows Angel to have more strategic options.

For example, the $k=1$ case corresponds to the three-state quantum walk. In the most widely studied quantum walk, a walker can only travel to an adjacent point in one step, but in the quantum angel problem, Angel can travel anywhere within a distance $k$. Therefore the transition equation of Angel is  
\begin{equation}
    \ket{\Psi_{n+1}(x)}=\sum_{y=x-k}^{x+k}U_{x-y}\ket{\Psi_{n}(y)},~x,y\in\mathbb{Z}.
\end{equation}
We put the initial state as $\ket{\Psi_{0}(0)}={}^T[a_{-k},\cdots, a_k]$ satisfying
\begin{equation*}
    \sum_{l=-k}^k|a_l|^2=1. 
\end{equation*}
The general state of Angel at $t=n$ is given by 
\begin{equation}
    \Psi(t=n)={}^T[\cdots, \ket{\Psi_{n}(-1)}, \ket{\Psi_{n}(0)}, \ket{\Psi_{n}(1)},\cdots].
\end{equation}

Let us give an alternative definition of Angel living in $H_p\otimes H_k$, where $l^2(\mathbb{Z})$ is the position space and $H_k$ is the space of Angel's power. The transition operator of Angel with power $k$ is 
\begin{equation}
    S_m=\sum_{x\in\mathbb{Z}}\ket{x+m}\bra{x}, m=-k,\cdots, k. 
\end{equation}
The dynamics of Angel is defined as follows
\begin{equation}
    U_t=\left(\sum_{m=-k}^kS_m\otimes U_m\right) (I_\infty\otimes U(t)),
\end{equation}
where $I_\infty$ is the identity operator and $U(t)$ is a strategy of Angel at time $t$. Then a general state of Angel after $n$ time steps is given as 
\begin{equation}
   \Psi(t=n)= \left(\bigotimes_{t=1}^n U_t\right)\Psi(t=0).
\end{equation}

\subsection{Quantum Angel v.s. Devil}
Let us now discuss variants of the quantum angel problem. As already mentioned several times, the main reasons why the quantum angel problem differs from the traditional angel problem are that 1) Angel is a quantum walker and 2) it is possible to impose limits on the quantum resources available to Angel and Devil. 

The first point that Angel is a quantum walker gives good reason why this game could be interesting even if it is one-dimensional. As is already widely known, the probability distribution of a one-dimensional quantum walk is quite different from that of a classical random walk~\cite{10.1145/380752.380757,2001quant.ph..3020C,konno2002quantum}. Especially in the case of the Hadamard walk, the probability distribution of the quantum walker spreads linearly with time. This means that it will be extremely difficult for Devil to predict Angel's location. These are all for Angel with power $k=1$, i.e., when Angel can only move to adjacent sites in one time step, but it would be even more difficult to predict the quantum walker's location when $k$ is 3 or greater, as we formulate in this paper. Our method can be easily extended to quantum walks on general graphs~\cite{2000quant.ph.12090A}.

Regarding the second point, there are actually a number of interesting variations. From the practical point of view of currently available quantum computers, games using non-universal quantum computers are very interesting. Research on quantum computers, which are not universal but are more powerful than classical computers, is very useful and important today~\cite{2002PhRvA..65c2325T,doi:10.1098/rspa.2010.0301,PhysRevLett.81.5672}. They are the most important class of problems, not only in the study of the polynomial hierarchy of computational complexity, but also for potential applications of NISQ devices. Depending on the quantum resources available to Angel and Devil, the game is divided into the following nine cases (Table \ref{tab:my_label}).  
\begin{itemize}
    \item Both Angel and Devil can use unlimited quantum resources $\ctext{1}$. 
    \item Angel can use better quantum resources than Devil $\ctext{2},\ctext{3},\ctext{6}$. 
    \item Devil can use better quantum resources than Angel $\ctext{4},\ctext{7},\ctext{8}$. 
    \item Both Angel and Devil can use non-universal quantum computers $\ctext{5}$. 
    \item Both Angel and Devil can only perform classical calculations $\ctext{9}$. 
\end{itemize}

It is interesting to ask whether Angel or Devil have the advantage or the winning strategy in these nine cases. These are new questions that have not been asked in the traditional angel problem. In the classical case, Angel has no chance of winning against Devil in one dimension. So if Angel capable of universal quantum computation has a high probability of beating Devil, then by properly limiting the resources of Devil and Angel, it is always possible to create an environment in which both have a 50\% chance of winning. 

So far we have considered limiting the ability of players, but we can also extend it. Studying the behavior of players with computational capabilities beyond quantum theory is also useful in understanding the upper bounds of quantum computation. Indeed, a number of such classes of theories have been investigated in the field of quantum computation~\cite{2004quant.ph..1062A,aaronson2009closed}. For example, if Angel is capable of post-selection, the desired result can be obtained with probability 1, in a way that there is no probability fluctuation. It would be an interesting question of mechanism design to investigate how much to limit or extend the computational power of Angel and Devil to make a good game.

\begin{table}[H]
    \centering
    \begin{tabular}{c||c|c|c}
       Angel $\backslash$ Devil  & Universal Quantum & Non-Universal & Classical\\
       \hline \hline
      Universal Quantum & \ctext{1}  & \ctext{2} & \ctext{3}\\\hline
      Non-Universal & \ctext{4} & \ctext{5} &\ctext{6}\\\hline
      Classical & \ctext{7} & \ctext{8} & \ctext{9}\\
    \end{tabular}
    \caption{Classification of quantum angel problem based on quantum resources available to Angel and Devil.}
    \label{tab:table}
\end{table}

\section{Quantum Circuit of Quantum Angel Problem}
\subsection{Operators of Devil}
Here we create operators of Devil for placing a block at $x$. For this Devil needs to prepare two operators: creating a block and diagnosing Angel's occupancy. The second operation is needed since Devil is allowed to place a block if Angel is not detected there. We first explain how to create the operator to create a block. 

\begin{figure}[H]
    \centering
\begin{quantikz}
\gategroup[wires=2,steps=4,style={rounded corners,fill=blue!10}, background]{}
\ket{\psi_x}   & \ctrl{1} &\targ{} &\ket{1}\qw\\
\ket{0}& \targ{} & \octrl{-1}& \meter{}
\end{quantikz}
    \caption{Creating a block at $x$.}
    \label{fig:block}
\end{figure}
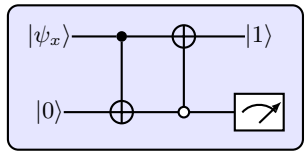

We denote the states without and with blocks by $\ket{0}$ and $\ket{1}$, respectively. In general, the block's state at $x$ can be a superposition
\begin{equation}
    \ket{\psi_x}=\alpha_x\ket{0}+\beta_x\ket{1}
\end{equation}
and the state of all blocks looks like 
\begin{equation}
    \ket{\psi}={}^T\left[\cdots,\binom{\alpha_{-1}}{\beta_{-1}},\binom{\alpha_0}{\beta_0},\binom{\alpha_{+1}}{\beta_{+1}},\cdots\right]. 
\end{equation}
Devil can place a block at $x$ by using the following operators. Devil first adds $\ket{0}$ to each $\ket{\psi_x}$ and operates $\ket{0}\bra{0}\otimes I+\ket{1}\bra{1}\otimes X$ to $\ket{\psi_x}\ket{0}$. By operating $X\otimes \ket{0}\bra{0}+I\otimes \ket{1}\bra{1}$ to it again and measuring the second qubit, Devil can obtain $\ket{1}$ with probability 1. This procedure is summarized in the following diagram and circuit (Fig.\ref{fig:block})
\begin{equation}
    (\alpha_x\ket{0}+\beta_x\ket{1})\ket{0}\xrightarrow{\text{CNOT}} \alpha_x\ket{0}\ket{0}+\beta_x\ket{1}\ket{1}\xrightarrow{\text{CNOT}} \alpha_x\ket{1}\ket{0}+\beta_x\ket{1}\ket{1}\xrightarrow{\text{Measurement}} \ket{1}
\end{equation}

\begin{figure}[H]
    \centering
\begin{quantikz}
\gategroup[wires=2,steps=3,style={rounded corners,fill=blue!10}, background]{}
\ket{\varphi_x}   & \ctrl{1} & \qw \\
\ket{0} & \targ{} & \meter{}
\end{quantikz}
    \caption{Diagnosing Angel's presence at $x$.}
    \label{fig:diagnose}
\end{figure}
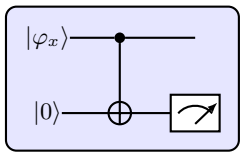

Before placing a block, Devil needs to know whether or not Angel is present at that location. For this, Devil adds a qubit on each $x$ and denotes $\ket{1}$ if there is Angel at x and $\ket{0}$ if there is not, respectively. Then whether Angel is at $x$ or not can generally be expressed in the following state.
\begin{equation}
    \ket{\varphi_x}=a_x\ket{0}+b_x\ket{1},
\end{equation}
which should be normalized $|a_x|^2+|b_x|^2=1$. In general Devil can diagnose the presence of Angel by the following state
\begin{equation}
    \ket{\varphi}={}^T\left[\cdots,\binom{a_{-1}}{b_{-1}},\binom{a_0}{b_0},\binom{a_{+1}}{b_{+1}},\cdots\right]. 
\end{equation}
Then Devil adds another $\ket{0}$ to each $\ket{\varphi_x}$ and applies the CNOT operation
\begin{equation}
    (a_x\ket{0}+b_x\ket{1})\ket{0}\xrightarrow{\text{CNOT}} a_x\ket{0}\ket{0}+b_x\ket{1}\ket{1}
\end{equation}
If the measurement result of the second qubit is 0, we say that Angel is not at $x$. If it is 1, we say that Angel is there. Clearly one can get $\ket{0}$ and $\ket{1}$ with probability $|a_x|^2$ and $|b_x|^2$, respectively. The circuit of this operation is drawn in Fig.\ref{fig:diagnose}

\subsection{Operations of Angel}
Here we define the operator to determine where Angel can move. It must be consistent with the rules of the game that Angel cannot move to a point where a block is placed. Recall that the block's state at $x$ is represented by
\begin{equation}
    \ket{\psi_x}=\alpha_x\ket{0}+\beta_x\ket{1}, 
\end{equation}
Angel can know if there is a block there by applying the CNOT operation. 
\begin{equation}
    (\alpha_x\ket{0}+\beta_x\ket{1})\ket{0}\xrightarrow{\text{CNOT}}\alpha_x\ket{0}\ket{0}+\beta_x\ket{1}\ket{1}
\end{equation}
By measuring the second qubit, Angel finds a block with probability $|\beta_x|^2$. 

\section{Conclusion and Future Directions}
In this study, we formulated the concept of quantum extensive-form games mathematically. We also confirmed that several important quantum algorithms, such as the Grover algorithm and quantum GAN, can be expressed in quantum extensive form. Given that a number of important models in game theory are described in extensive forms, it is surprising that quantum extensive-form games have received so little attention. Until now, an excessive amount of attention has been paid to the EWL protocol in quantum games (single stage games in most cases), but we hope that this research will be one of the catalysts for turning attention to the study of quantum extensive-form games.  

Any number of research topics could follow this paper. Pursuing Nash equilibrium solutions of quantum extensive-form games is a natural and important mathematical topic, and contributes significantly to the development of models in economics. It would also be very meaningful to discuss existing concrete non-cooperative quantum games in a quantum extensive form in order to discuss repeated games. In existing quantum games, single-stage games have been studied extensively, while the importance of repeated games has not been recognized. As mentioned many times in this paper, in order to take full advantage of quantum nature in game theory, it is wiser to use the pairwise annihilation of branches which occurs through quantum mechanical transitions of states, rather than considering entanglement alone. The excessive focus on entanglement in the existing quantum game does not necessarily match the study of algorithms that enable quantum advantage. In fact, it is known that universal quantum computation can be achieved with only a small amount of entanglement~\cite{PhysRevLett.110.060504}. This could be the background to the fact that existing quantum games have yielded few useful results, but as this study has shown, the situation could be improved by considering quantum extensive-form games.

Regarding the quantum angel problem, a number of interesting studies would be possible. First, it would be intriguing to examine to which class of the polynomial hierarchy the nine cases of games presented in Table \ref{tab:table} belong, and whether Angel or Devil are favored in each case. First of all, determining whether or not Angel is alive at a given time $t$ is a decision problem. It is also a decision problem to look at the game phase at each time and determine whether Devil or Angel has the advantage. This could actually be formulated in a meaningful way as a measurement-based quantum computation (MBQC) problem. MBQC is a way of universal computation which uses  measurement to achieve unitary time evolution of circuits, by measuring qubits one-by-one~\cite{briegel2009measurement,wei2021measurement}. In Quantum Angel Problem, Devil measures eqch qubit sequentially to see where Angel's position (Fig~\ref{fig:diagnose}). With  Devil measuring qubits each time, the time evolution of the 1d chain of qubits is equivalent to generating a unitary gate in MBQC.

\begin{figure}[H]
    \centering
    \includegraphics[width=8cm]{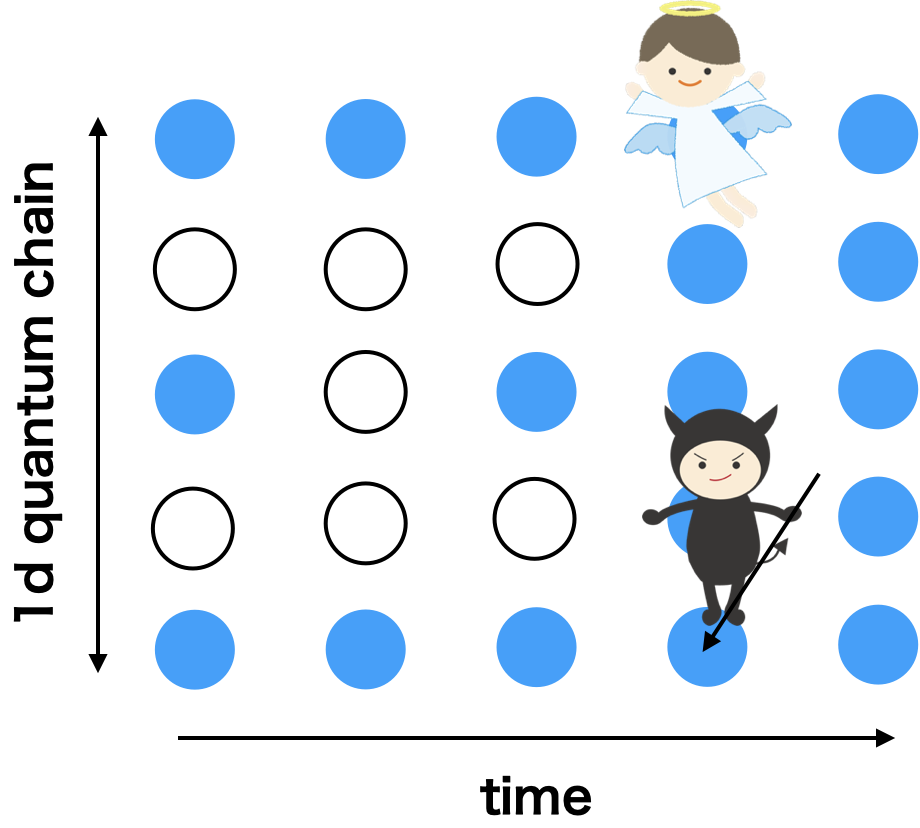}
    \caption{Schematic picture of a correspondence between Quantum Angel Problem and measurement based quantum computation (MBQC).}
    \label{fig:MBQC}
\end{figure}

To make this game as powerful as universal quantum computation, it is necessary for Devil to be able to measure multiple qubits at each round of the game (Fig.~\ref{fig:MBQC}). As Life Game invented by Conway in 1970 is as powerful as universal classical computation~\cite{gardner1970mathematical,berlekamp2004winning}, we expect the most general form of Quantum Angel Problem is a way of doing universal quantum computation.
\begin{conjecture}
    Quantum Angel Problem where Devil is allowed to measure more than 1 qubit at each time step of the game is as powerful as universal quantum computation.
\end{conjecture}

It would be fun to actually implement and play this game on a NISQ device.

\section*{Data Availability}
Non-digital data supporting this study are curated. Data sharing is not applicable to this article as no new data were created or analyzed in this study.  

\section*{Acknowledgement}
The author is supported by PIMS Postdoctoral Fellowship Award. The author was also supported by the U.S. Department of Energy, Office of Science, National Quantum Information Science Research Centers, Co-design Center for Quantum Advantage (C2QA) under Contract No.DESC0012704. The author thanks Dmitri Kharzeev, Adam Lowe, Steven Rayan, Hiroki Sukeno and Tzu-Chieh Wei for useful discussion, collaborations and encouragement. The author is grateful to Megumi Ikeda for providing him with the cartoons. 

\section*{References}
\bibliographystyle{utphys}
\bibliography{ref}
\end{document}